\documentclass[aps,prl,numerical,superscriptaddress,showpacs,floatfix,nofootinbib,reprint]{revtex4-1}
\usepackage{amsmath,amssymb,mathrsfs,bm}
\usepackage{url}
\usepackage{hyperref}
\hypersetup{
colorlinks=true,
linkcolor=blue,
anchorcolor =red,
citecolor=blue,
filecolor = red,
urlcolor=blue,
pdfauthor=author}
\usepackage{txfonts}
\usepackage[utf8]{inputenc}
\usepackage{amsmath}
\usepackage{graphicx}
\def\av<#1>{\left\langle\,#1\,\right\rangle}
\def\ev<#1>{\left\langle\,#1\,\right\rangle_{\rm{ev}}}

\begin{document}


\title{Anisotropy Scaling Functions in Heavy-Ion Collisions: Insights into the `Ultra-Central Flow Puzzle' and Constraints on Transport Coefficients and Nuclear Deformation}

\author{ Roy~A.~Lacey}
\email[E-mail: ]{Roy.Lacey@Stonybrook.edu}
\affiliation{Department of Chemistry, 
Stony Brook University, \\
Stony Brook, NY, 11794-3400, USA}
%
%

\date{\today}

\begin{abstract}
Anisotropy scaling functions derived from comprehensive measurements of transverse momentum- and centrality-dependent anisotropy coefficients \(v_2(p_T,\text{cent})\) and \(v_3(p_T,\text{cent})\) in Pb+Pb collisions at 5.02 and 2.76 TeV, Xe+Xe collisions at 5.44 TeV and Au+Au collisions at 0.2 TeV, offer new insights into the `ultra-central flow puzzle'. These functions integrate diverse measurements into a single curve, clarifying anisotropy attenuation throughout the entire \(p_T\) and centrality range. They reveal the influence of initial-state eccentricities (\(\varepsilon_{n}\)), dimensionless size (\(\mathbb{R}\)), radial flow, viscous correction to the thermal distribution function (\(\delta_f\)), the medium's stopping power (\(\Hat{q}\)), and specific shear viscosity (\(\eta/s\)) on the observed anisotropies. This analysis not only enhances understanding of transport coefficients but also provides crucial constraints on nuclear deformation.
\end{abstract}


\maketitle

Azimuthal anisotropy measurements are pivotal in the study of the quark-gluon plasma (QGP) produced in heavy ion collisions at the Relativistic Heavy Ion Collider (RHIC) and the Large Hadron Collider (LHC). These measurements, crucial for determining the temperature ($T$) and baryon chemical potential ($\mu_B$) dependence of transport coefficients, are quantified by the complex coefficients \cite{Bilandzic:2010jr,Luzum:2011mm,Teaney:2012ke}:
\begin{equation}
 V_n \equiv v_ne^{in\Psi_n} = \langle e^{in\phi} \rangle,
\label{Vndef}
\end{equation}
where $v_n$ represents the degree of azimuthal anisotropy, $\Psi_n$ is the event-plane angle, and $\langle \cdot \rangle$ denotes averaging over the single-particle spectrum within an event. 

These \(v_n\) values are linked to the Fourier coefficients \(v_{nn}\), which describe the intensity of two-particle correlations in azimuthal angle differences \(\Delta\phi = \phi_{\mathrm{i}} - \phi_{\mathrm{j}}\) \cite{Poskanzer:1998yz,Lacey:2005qq}:
\begin{eqnarray}
\frac{dN^{pairs}}{d\Delta\phi} \propto 1 + 2\sum_{n=1}^{\infty} v_{nn} \cos(n\Delta\phi), \nonumber \\
v_{nn}(p_T^{i},p_T^{j}) = v_n(p_T^{i})v_n(p_T^{j}) + \delta_{\text{NF}},
\label{eq:2}
\end{eqnarray}
where \(\delta_{\text{NF}}\) indicates non-flow contributions, minimized through specific experimental methodologies \cite{Lacey:2005qq,Luzum:2010fb,Retinskaya:2012ky,ATLAS:2012at}.

The coefficients \(v_n\) are fundamentally linked to collective flow dynamics for \(p_T \alt 4-5\) GeV, transitioning to jet quenching at higher \(p_T\). This behavior is supported by extensive research~\cite{Song:2010mg,Alver:2008zza,Alver:2010rt,Ollitrault:2009ie,Dusling:2009df,Lacey:2010fe,Shen:2011eg,Niemi:2012aj,Fu:2015wba,Andres:2015ara,STAR:2022gki,STAR:2018fpo,ALICE:2016kpq,Qiu:2011iv,Adare:2011tg,Magdy:2018itt,Adamczyk:2016gfs,STAR:2015rxv,Adamczyk:2015obl,Adamczyk:2016exq,Adam:2019woz,Gardim:2014tya,Holopainen:2010gz,Qin:2010pf,Qiu:2011iv,Gale:2012rq,JET:2013cls,Liu:2018hjh,PHENIX:2001hpc,STAR:2002ggv,Zhang:2008fh,ALICE:2013dpt,Mehtar-Tani:2013pia,Qin:2015srf}, which highlights the impact of radial expansion, \(v_n\)-fluctuations, \(p_T\)-dependent viscous attenuation, jet quenching, and initial-state anisotropy on \(v_n\). The transverse plane density profile \(\rho(r,\varphi)\) is quantified by complex eccentricity coefficients:
\begin{equation} 
{\mathcal{E}_n \equiv \varepsilon_ne^{in\Phi_n} =
\frac{\int d^2r_\perp\, r^m\,e^{in\varphi}\, \rho(r,\varphi)} {\int d^2r_\perp\, r^m\,\rho(r,\varphi)}}, 
\label{enpc}
\end{equation}
where $r$ and $\varphi$ represent the radius and azimuthal angle, respectively, and ${\Phi_n}$ denotes the angle of the ${n^{\rm th}}$-order participant plane (${m=n}$ for ${n{\geq}2}$ and ${m=3}$ for ${n=1}$) \cite{Qiu:2011iv, Fu:2015wba, Niemi:2015qia, Noronha-Hostler:2015dbi}. The deformed Woods-Saxon distribution, employed to model nucleon configurations in non-spherical nuclei, is described by:
\begin{align}
    \rho(r,\theta, \varphi) &= \frac{\rho_0}{1 + \exp \left(\frac{r - R(\theta, \varphi)}{a}\right)},  \label{enpc1} \\
		R(\theta,\varphi) &= R_0 \left\{ 1 + \beta_2 \left[  \cos \gamma Y_{20}(\theta,\varphi) + 
		\sin \gamma Y_{22}(\theta,\varphi) \right] \right\} \nonumber, 
\end{align}
where $\rho_0$ is the central density, $R_0$ the nuclear radius, and $a$ the skin thickness. The spherical harmonics $Y_{lm}$, along with coefficients $\beta_2$ and $\gamma$, shape the nucleus, with $\beta_2$ measuring deformation magnitude and $\gamma$ indicating asymmetry degrees between 0 and \(60^\circ\). Fluctuations in the initial state density profile lead to variations in $\varepsilon_n$.

Dynamical models based on relativistic hydrodynamics, which suggest a roughly linear relationship \(v_n \propto \varepsilon_n\) \cite{Noronha-Hostler:2015dbi, Sievert:2019zjr, Rao:2019vgy}, have successfully replicated the observed magnitudes and trends of \(v_n\) coefficients \cite{Qin:2010pf, Schenke:2011tv, Qiu:2011iv, Shen:2011eg, Bozek:2011ua, Gardim:2012yp, Hirano:2012kj}. These models play a central role in efforts employing Bayesian inference to constrain the transport properties of the QGP \cite{Bernhard:2016tnd, Bernhard:2019bmu, Moreland:2018gsh, JETSCAPE:2020shq, JETSCAPE:2020mzn, Nijs:2020roc, Auvinen:2020mpc, Parkkila:2021tqq}. However, they encounter significant challenges in ultra-central collisions (cent \(\alt 1\%\)), where they struggle to simultaneously predict \(v_2\) and \(v_3\) accurately \cite{Giannini:2022bkn}. Typically, these models either overestimate \(v_2\), underestimate \(v_3\), or display both discrepancies collectively termed the `ultra-central flow puzzle.'

The inability of hydrodynamic models to accurately describe both \(v_2\) and \(v_3\) in ultra-central collisions presents a significant challenge, contradicting the expectations set for these collisions. Conventionally, models were expected to perform well in central collisions due to higher charge particle multiplicities \({\left\langle {{\text{N}}_{\text{chg}}} \right\rangle}\) and larger volumes of locally thermalized domains. This discrepancy raises doubts about whether these widely used models are missing crucial components necessary for accurately depicting the initial state and transport coefficients.

Despite extensive efforts to resolve the `ultra-central flow puzzle' by refining initial conditions \cite{Luzum:2012wu, Denicol:2014ywa, Shen:2015qta, Bhalerao:2015iya, Loizides:2016djv, Giacalone:2019kgg, Gelis:2019vzt, Carzon:2020xwp, Snyder:2020rdy, Zakharov:2020irp}, examining transport coefficients \cite{Luzum:2012wu, Rose:2014fba, Shen:2015qta, Plumari:2015cfa}, and evaluating equations of state \cite{Alba:2017hhe}, a definitive resolution remains elusive. The ongoing investigations highlight the complexity of accurately modeling these critical aspects of QGP behavior.

In this study, an anisotropy scaling function is developed based on the concept that diverse measurements of \(v_2(p_T,\text{cent})\) and \(v_3(p_T,\text{cent})\) can be unified into a single, coherent function. This scaling function consolidates various parameters that affect \(v_n(p_T,\text{cent})\), such as initial-state eccentricities (\(\varepsilon_{n}\)), dimensionless size (\(\mathbb{R} \propto RT\)), radial flow magnitude, the medium's stopping power (\(\hat{q}\)), and the specific shear viscosity or viscosity-to-entropy ratio (\(\eta/s \propto T^3/\hat{q}\)), along with the viscous correction to the thermal distribution function (\(\delta_f\))~\cite{Majumder:2007zh, Dusling:2009df}.

These parameters contribute to the expression of the anisotropy coefficients derived from the dispersion relation for sound propagation. This relationship is described by Eq.~\ref{vnpc}~\cite{Staig:2010pn, Gubser:2010ui, Lacey:2013is, Liu:2018hjh}:
\begin{equation}
v_n(p_T,\text{cent}) = \varepsilon_n(\text{cent}) e^{-\frac{\beta}{\mathbb{R}} \left[n(n + \kappa p_T^2)\right]}, \quad n=2,3,
\label{vnpc}
\end{equation}
where \(\beta \propto \eta/s\), \(\delta_f = \kappa p_T^2\)~\cite{Dusling:2009df, Liu:2018hjh}, and \(\mathbb{R} \propto \langle N_{\text{chg}} \rangle_{|\eta|\leq 0.5}^{1/3}\) relates to the mid-rapidity (\(|\eta| \leq 0.5\)) charged particle multiplicity. The transition from flow to jet quenching at higher \(p_T\) is managed by maintaining consistency between \(\eta/s\) and \(\hat{q}\), ensuring a smooth transition in \(\delta_f\) across low and high momentum regions. This shift is facilitated by fixing the \(\kappa p_T^2\) term in Eq.~\ref{vnpc} to remain constant for \(p_T\) values above approximately 4.5~GeV/c, marking the threshold between flow-dominated and jet-quenching domains.

\begin{figure}[tbh]
    \centering
    \includegraphics[clip,width=1.0\linewidth]{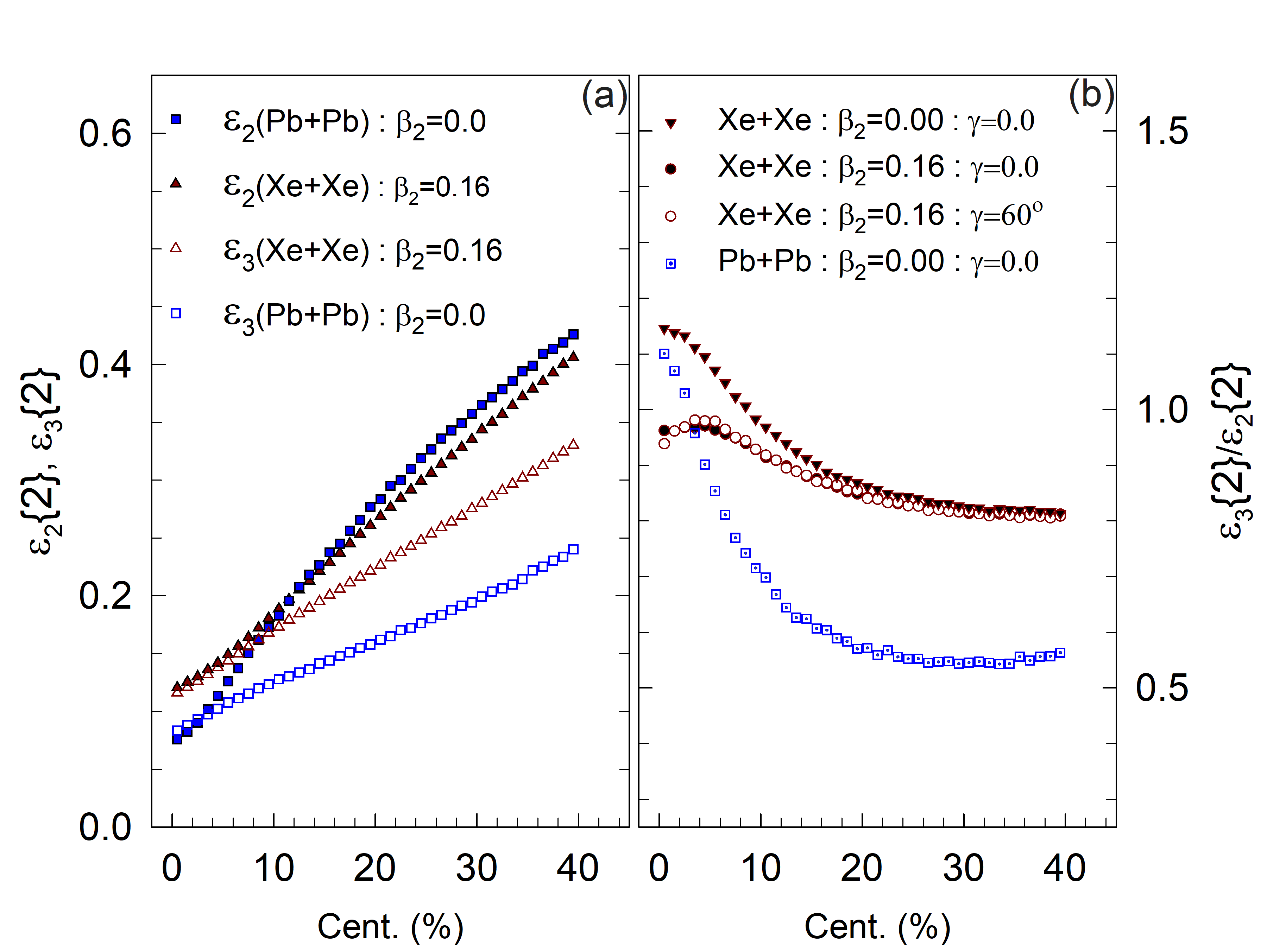}
	\vskip -0.3 cm
    \caption{(Color Online)
    Panel (a) contrasts the centrality-dependent values of \(\varepsilon_2(\text{cent})\) and \(\varepsilon_3(\text{cent})\) for 5.02~TeV Pb+Pb collisions featuring spherical Pb nuclei and 5.44~TeV Xe+Xe collisions for deformed Xe nuclei. In panel (b), the ratios \(\varepsilon_3(\text{cent})/\varepsilon_2(\text{cent})\) are compared for spherical Pb nuclei (\(\beta_2 = 0\)) and both spherical and deformed Xe nuclei, as indicated.}
    \label{fig1}
\end{figure}

Equation~\ref{vnpc} provides crucial insight into the behavior of anisotropy coefficients in the domain of the most central collisions, identified by the peak value \(\mathbb{R}_0\). It establishes a scaling relationship that correlates the harmonic \(v_n(p_T, 0)\) measured in ultra-central events with \(v_n'(p_T, \text{cent})\) at varying centralities, each linked to a specific \(\mathbb{R}'\) value:
\begin{equation}
\frac{v_n(p_T,0)}{\varepsilon_n(0)} e^{\frac{n \beta}{\mathbb{R}_0} [n + \kappa p_T^2]} = \frac{v_n'(p_T,\text{cent})}{\varepsilon_n'(\text{cent})} e^{\frac{n \beta}{\mathbb{R}_0} [n + \kappa p_T^2] \left(\frac{\mathbb{R}_0}{\mathbb{R}'} - 1\right)}.
\label{vnpc1}
\end{equation}
This equation highlights the attenuation of ${\varepsilon_n}$-scaled anisotropy in ultra-central collisions and its relative change at other centralities. Fig.~\ref{fig1} displays the centrality-dependent values of \(\varepsilon_2(\text{cent})\) and \(\varepsilon_3(\text{cent})\) for 5.02~TeV Pb+Pb and 5.44~TeV Xe+Xe collisions, elucidating the effects of nucleus size and deformation.
\begin{figure*}[tbh]
    \centering
    \includegraphics[clip,width=0.80\linewidth]{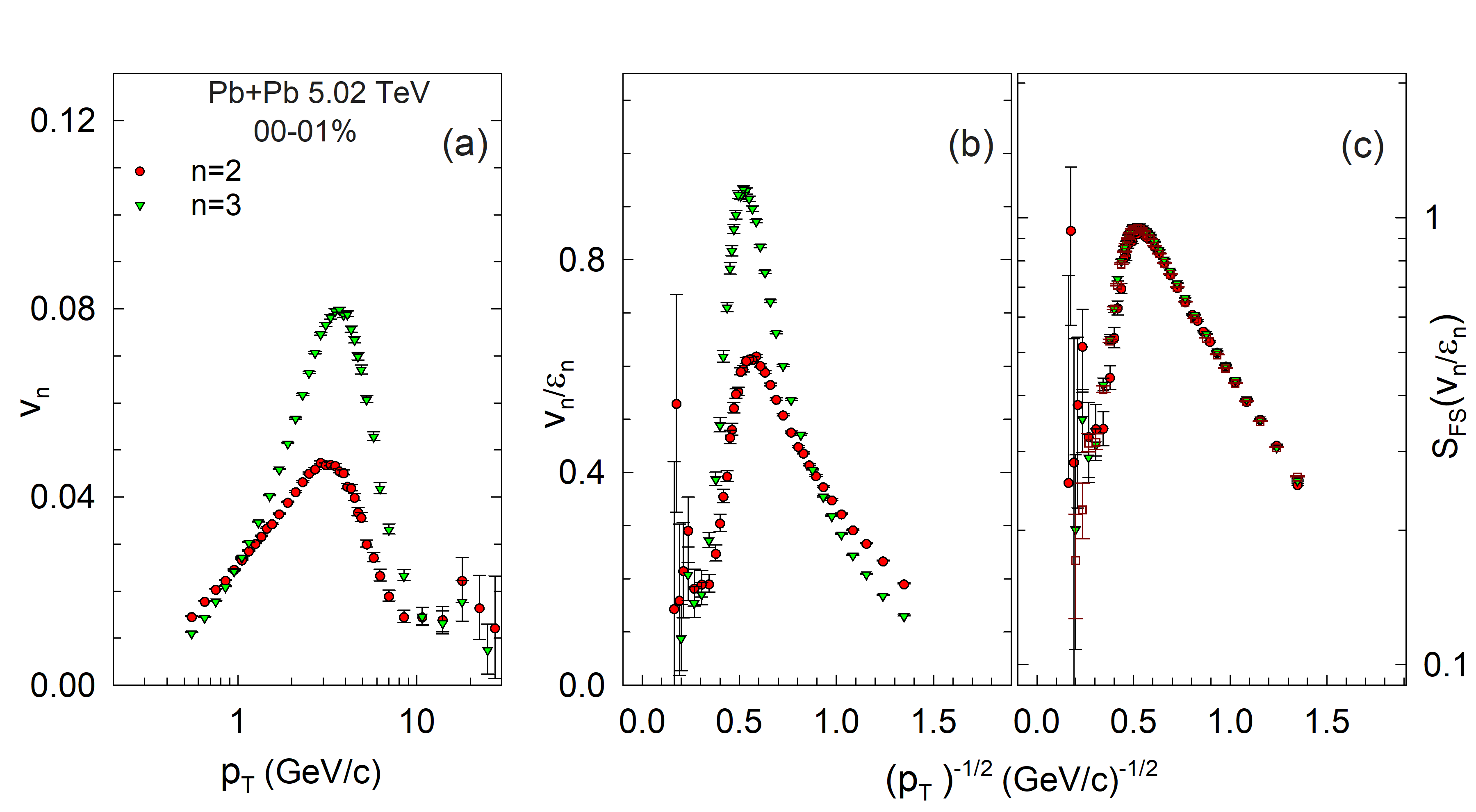}
		\vskip -0.3 cm
    \caption{(Color Online)
		 Comparison of \(v_2(p_T)\) and \(v_3(p_T)\) in panel (a), their eccentricity-scaled values \(v_2/\varepsilon_2\) 
		and \(v_3/\varepsilon_3\) in panel (b), and the resulting scaling function in panel (c) for 0.1\% central Pb+Pb collisions 
		at 5.02 TeV. Panel (c) also includes scaled results for 30-40\% central Pb+Pb collisions. The data are sourced 
		from the ATLAS collaboration \cite{ATLAS:2018ezv}.
            }
    \label{fig2}
\end{figure*}

Within a fixed centrality for a given system, a significant correlation exists between \(v_2(p_T, \text{cent})\) and \(v_3(p_T, \text{cent})\), both of which are influenced by the same underlying factors: radial flow, \(\eta/s\), and \(\hat{q}\). This correlation is expressed as:
\begin{equation} 
\frac{v_2(p_T,\text{cent})}{\varepsilon_2(\text{cent})} e^{\frac{2\alpha \beta}{\mathbb{R}_0}} = \left(\frac{v_3(p_T,\text{cent})}{\varepsilon_3(\text{cent})}\right)^{\frac{2}{3}},
\label{vnpc2}
\end{equation}
where \(\alpha\) is a system-dependent but centrality-independent normalizing constant. 

The radial flow difference for a given system, relative to Pb+Pb collisions at 5.02 TeV at the same centrality, can be scaled via the expression:
\begin{equation} 
\frac{v_n'(p_T)}{\varepsilon_n'} e^{\frac{n \beta}{\mathbb{R}_0} [n + \kappa p_T^2] \left(\frac{\mathbb{R}_0}{\mathbb{R}'} - 1\right)} = \left(\frac{v_n''(p_T)}{\varepsilon_n''}\right)^{(1-\zeta)} e^{\frac{n \beta''(1-\zeta)}{\mathbb{R}_0} [n + \kappa p_T^2] \left(\frac{\mathbb{R}_0}{\mathbb{R}'} - 1\right)},
\label{vnpc3}
\end{equation}
where \(\zeta\) parametrizes the difference in radial flow.

Equations~\ref{vnpc1}, \ref{vnpc2}, and \ref{vnpc3} encapsulate the intricate dependencies of anisotropy coefficients across the spectrum from ultra-central to peripheral collisions for various systems and energies. These relationships suggest that \(v_2(p_T, \text{cent})\) and \(v_3(p_T, \text{cent})\) measurements can be scaled to converge onto a single scaling function \(S_{\!\rm{FS}}\). Identifying such a scaling function would provide robust evidence for the coherence of the scaling coefficients and validate the reliability of the corresponding eccentricity spectrum and its ratios, offering deeper insight into the underlying collision dynamics and transport coefficients.

As illustrated in Fig.~\ref{fig1}, the eccentricity ratio plays a pivotal role in deciphering the nuanced effects of centrality and deformation on collision dynamics. The calculated values of \(\varepsilon_2(\text{cent})\), \(\varepsilon_3(\text{cent})\), and their ratios for Pb+Pb and Xe+Xe collisions at 5.02~TeV and 5.44~TeV, respectively, reveal distinct dependencies that significantly refine the eccentricity spectrum. This refinement is crucial for illuminating the initial-state deformation of the Xe nucleus, which in turn affects the entire evolution of the collision. The findings underscore that measurements in ultra-central and central collisions are indispensable for accurately quantifying nuclear deformation, as these regions provide the most sensitive probes of the initial geometric asymmetries. Moreover, the observed insensitivity of the ratio \(\varepsilon_2(\text{cent})/\varepsilon_3(\text{cent})\) to variations in the parameter \(\gamma\) highlights its robustness across different asymmetries, making it a reliable indicator of deformation effects.

The data utilized in this study are sourced from the ATLAS~\cite{ATLAS:2012at,ATLAS:2018ezv}, ALICE~\cite{ALICE:2018lao,ALICE:2018rtz,ALICE:2018yph}, and PHENIX~\cite{PHENIX:2014uik,PHENIX:2010nlr} collaborations, which encompass \(v_{2}(p_T, \text{cent})\) and \(v_{3}(p_T, \text{cent})\) measurements for Pb+Pb collisions at \(\sqrt{s_{NN}} = 2.76\) and 5.02 TeV, Xe+Xe collisions at 5.44 TeV, and Au+Au collisions at 0.2 TeV. The centrality-dependent \(\left\langle {\text{N}_{\text{chg}}} \right\rangle_{|\eta|\le 0.5}\) values required for this study were derived from corresponding multiplicity density measurements~\cite{ALICE:2010mlf,ALICE:2015juo,ALICE:2018cpu,CMS:2019gzk,Lacey:2016hqy}. The previously established value of \(\kappa = 0.17\) (GeV/c)$^{-2}$~\cite{Liu:2018hjh} was adopted to compute \(\delta{f}\). The eccentricities were calculated according to the procedure outlined in Eq.~\ref{enpc}, using a Monte Carlo quark-Glauber model (MC-qGlauber) with fluctuating initial conditions~\cite{Liu:2018hjh}. This model, which is based on the widely utilized MC-Glauber model~\cite{Miller:2007ri,PHOBOS:2006dbo}, takes into account the finite size of the nucleon, the nucleon's wounding profile, the quark distribution within the nucleon, and quark cross sections that accurately reproduce the NN inelastic cross-section for the corresponding beam energies. Calculations were performed for Au and Pb nuclei, as well as for Xe nuclei with varying degrees of initial-state deformation characterized by different values for the \(\beta_2\) and \(\gamma\) parameters (cf. Eq.~\ref{enpc1}). A systematic uncertainty of 2-3\% was estimated for the eccentricities based on variations in the model parameters.
\begin{figure*}[tbh]
    \centering
    \includegraphics[clip,width=0.85\linewidth]{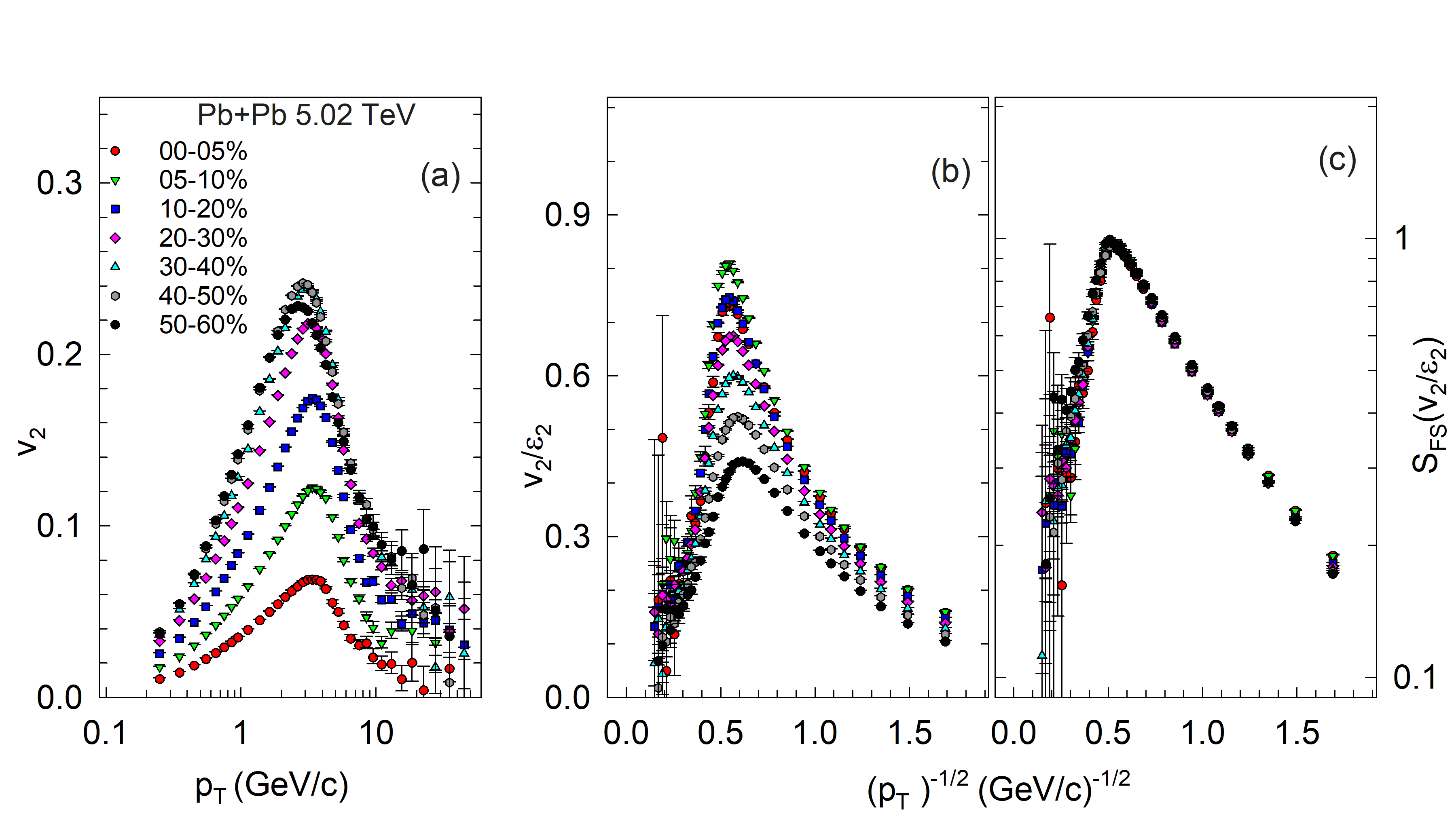}
		\vskip -0.3 cm
    \caption{(Color Online)
		Comparison of \(v_2(p_T,\rm{cent})\) in panel (a), their eccentricity-scaled values [\(v_2(p_T,\rm{cent})/\varepsilon_2(\text{cent})\)] in panel (b), and the resulting scaling function in panel (c) for Pb+Pb collisions at 5.02 TeV. Data sourced from the ALICE collaboration \cite{ALICE:2018rtz}.
            }
    \label{fig3}
\end{figure*}
\begin{figure*}[t]
    \centering
    \includegraphics[clip,width=0.85\linewidth]{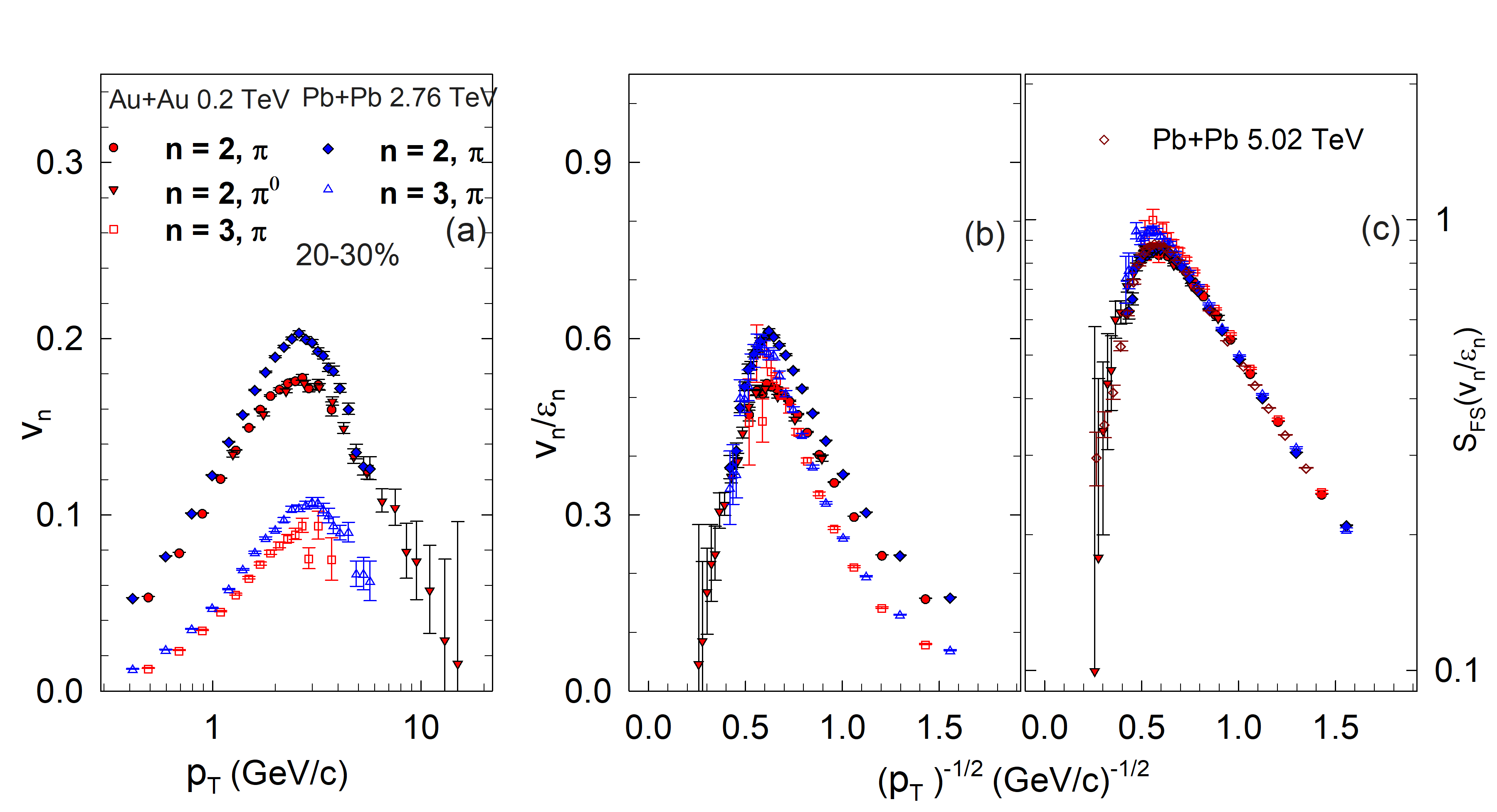}
		\vskip -0.3 cm
    \caption{(Color Online)
		 Comparison of the \(v_2(p_T)\) and \(v_3(p_T)\) values for charged ($\pi$) and neutral pions ($\pi^0$) in 0.2~TeV Au+Au collisions and charged pions in 2.7~TeV Pb+Pb collisions in panel (a), their eccentricity-scaled values \(v_2(p_T)/\varepsilon_2\) and \(v_3(p_T)/\varepsilon_3\) in panel (b), and the resulting scaling function in panel (c) for 20-30\% central collisions. Panel (c) also includes scaled results for pions in 5.02~TeV Pb+Pb collisions.The data are sourced from the PHENIX \cite{PHENIX:2014uik,PHENIX:2010nlr} and ALICE \cite{ALICE:2014wao,ALICE:2018yph} collaborations.
            }
    \label{fig4}
\end{figure*}
\begin{figure*}[hbt]
    \centering
    \includegraphics[clip,width=0.85\linewidth]{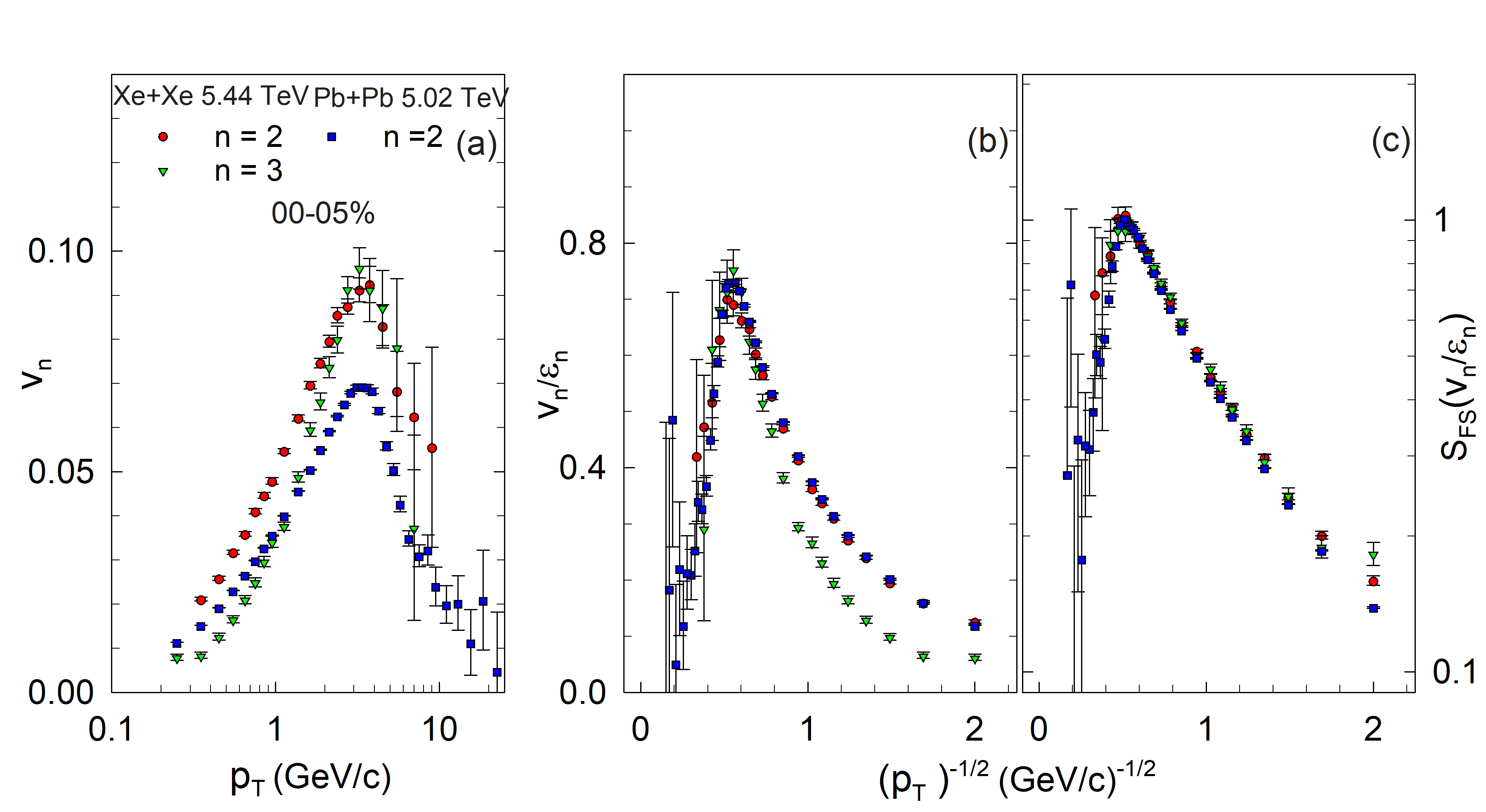}
		\vskip -0.3 cm
    \caption{(Color Online)
		 Comparison of the \(v_2(p_T,\text{cent})\) and \(v_3(p_T,\text{cent})\) values for Xe+Xe collisions and \(v_2(p_T,\text{cent})\) for Pb+Pb collisions in panel (a), their eccentricity-scaled values \(v_2(p_T,\text{cent})/\varepsilon_2(\text{cent})\) and \(v_3(p_T,\text{cent})/\varepsilon_3(\text{cent})\) in panel (b), and the resulting scaling function in panel (c) for 0-5\% central collisions. Panel (c) also includes scaled results for 30-40\% central Xe+Xe     collisions.The data are sourced from the ALICE collaboration \cite{ALICE:2018lao,ALICE:2018rtz} .
            }
    \label{fig5}
\end{figure*}

The scaling function is derived from differential measurements of \(v_{2}(p_T, \text{cent})\) and \(v_{3}(p_T, \text{cent})\) across a range of collision energies, including Pb+Pb collisions at 2.76 and 5.02 TeV, Xe+Xe collisions at 5.44 TeV, and Au+Au collisions at 0.2 TeV. This derivation utilized specific equations (\ref{vnpc1}, \ref{vnpc2}, \ref{vnpc3} and their combination), with the reference value \(\mathbb{R}_0\) for ultra-central Pb+Pb collisions at 5.02 TeV.

Figure~\ref{fig2} illustrates the scaling procedure for 0.1\% central Pb+Pb collisions at 5.02 TeV, employing \(1/\sqrt{p_T}\) on the x-axis in panels (b) and (c) to highlight the flow- and jet-quenching-dominated domains~\cite{Dokshitzer:2001zm,Lacey:2010fe}. Panel (a) emphasizes the discrepancy between \(v_2(p_T)\) and \(v_3(p_T)\), while panel (b) demonstrates that eccentricity scaling alone \([v_n(p_T)/\varepsilon_n]\) is insufficient to capture their difference. Panel (c) presents the resulting scaling function, showing a convergence of the data onto a single curve across both \(p_T\) domains.

Validation of the scaling function spanned the entire measurement range (0.0-0.1\% to 50-60\%) for specific parameters (\(\beta=0.88\) and \(\alpha=1\)), as demonstrated in Fig.~\ref{fig3} for \(v_{2}(p_T, \text{cent})\) measurements. The attenuation factor for the 0-5\% centrality cut reflects an average incorporating values from both ultra-central and non-ultra-central collisions, considering contributions from both scenarios. This validation provides robust evidence for the consistency of the scaling coefficients and the validity of the corresponding eccentricity spectrum and its ratios.

This consistency resolves the ultra-central flow puzzle by demonstrating that the derived scaling function accurately captures the behavior of \(v_2(p_T)\) and \(v_3(p_T)\) across different centralities, thus reconciling the discrepancies previously observed in ultra-central collisions.

Similarly, robust scaling functions, closely matching those for Pb+Pb collisions at 5.02 TeV, were obtained across the full range of measurements for Pb+Pb collisions at 2.76 TeV and Au+Au collisions at 0.2 TeV. Fig.~\ref{fig4} illustrates the scaling procedure and the resulting agreement between the scaled results obtained for pions in 20-30\% central Au+Au and Pb+Pb collisions. Notably, the consistency of the scaling functions across centrality for different beam energies suggests a negligible influence from non-flow effects. 

Characterized by scaling exponents \(\beta=0.84,\: \alpha=1.0,\: \zeta=0\) for Pb+Pb (2.76 TeV) and \(\beta=0.55,\: \alpha=1.6,\: \zeta=0.04\) for Au+Au (0.2 TeV), the scaling functions indicate (i) an approximate 5\% and 37\% reduction in \(\eta/s\) when the collision energy decreases from 5.02 TeV to 2.76 TeV and 0.2 TeV, respectively, and (ii) comparable radial flow for Pb+Pb collisions at 2.76 and 5.02 TeV, but comparatively smaller radial flow in Au+Au collisions

The observed scaling function, consistent across both flow and jet-quenching domains with similar scaling coefficients, provides strong constraints for \(\eta/s\) and \(\hat{q}\). It also supports the relationship between \(\eta/s\) and \(T^3/\hat{q}\) proposed by Majumder et al.~\cite{Majumder:2007zh,JETSCAPE:2020mzn}. This scaling, particularly the high \(p_T\) anisotropy's dependence on the dimensionless size \(\mathbb{R} \propto \left\langle {\text{N}_{\text{chg}}} \right\rangle^{1/3}\), underscores the influence of path length and highlights the role of radiative energy loss in shaping jet-quenching-induced anisotropy. 

Moreover, the seamless transition observed between the low and high momentum regions at the \(p_T\) threshold, which delineates flow from jet-quenching domains, provides crucial insights into both the absolute and relative magnitudes of \(\eta/s\) and \(T^3/\hat{q}\) \cite{Dusling:2009df}. This distinction is critical for determining whether the quark-gluon plasma (QGP) behaves as a strongly or weakly coupled system. A preliminary analysis, utilizing the \(\delta f\) formalism from Ref.~\cite{Dusling:2009df} and reconciling the values of \(\eta/s\) with \(\hat{q}\), suggests that \(\eta/s\) exceeds \(T^3/\hat{q}\). This relationship supports the interpretation of the QGP as a strongly coupled plasma, where shear viscosity is sufficiently low and the stopping power is moderate, leading to significant collective flow and jet-quenching effects \cite{JETSCAPE:2020mzn}.

The scaling function elucidates nuclear deformation, as illustrated in Figs.~\ref{fig1} and \ref{fig5}. Figure~\ref{fig5} compares \(v_{2}(p_T)\) and \(v_{3}(p_T)\) from Xe+Xe collisions against \(v_{2}(p_T)\) from Pb+Pb collisions at 0-5\% centrality, using eccentricities for the deformed Xe nucleus (\(\beta_2=0.16\)~\cite{Tsukada:2017llu}). Panel (a) highlights significant differences between these \(v_{n}(p_T)\) harmonics, with Xe+Xe showing larger magnitudes aligned with its higher eccentricities (cf. Fig.~\ref{fig1}). Panel (b) demonstrates that eccentricity scaling does not fully account for these differences but helps clarify the contrast in \(v_{2}(p_T)\) between Xe and Pb. Panel (c) illustrates the scaling function's capability to merge these data into a single curve across both flow- and jet-quenching-dominated regimes, validating the consistency of the scaling coefficients and the accuracy of the eccentricity spectrum for the deformed Xe nucleus. This scaling spans the centrality range of the available data (0-5\% to 40-50\%), with \(\beta=0.90\), \(\alpha=0.25\), \(\zeta=0.02\), and \(\beta_2=0.16\), suggesting (i) a mild deformation of the Xe nucleus, correlating with (ii) a comparatively smaller radial flow relative to that for Pb+Pb at 5.02 TeV, and (iii) an approximate 2\% increase in \(\eta/s\) when the collision energy increases from 5.02 to 5.44 TeV.

Applying scaling functions to analyze nuclear deformation significantly enhances the precision of our studies. This method (i) effectively compensates for anisotropy attenuation in ultra-central collisions, (ii) decouples initial- and final-state effects~\footnote{The correlation between final-state variables such as \(v_n\) and \(p_T\) is insufficient to accurately infer the initial state due to radial flow's significant and confounding influence, particularly in deformed systems.}, and (iii) relies on consistent measurements of \(v_2(p_T)\) and \(v_3(p_T)\) across events with identical \(\eta/s\), multiplicity, and radial flow characteristics. Consequently, it markedly reduces systematic uncertainties.

In summary, anisotropy scaling functions derived from comprehensive measurements of \(v_2(p_T,\text{cent})\) and \(v_3(p_T,\text{cent})\) in Pb+Pb collisions at 5.02 and 2.76 TeV, Xe+Xe collisions at 5.44 TeV and Au+Au collisions at 0.2 TeV, have been instrumental in unraveling the `ultra-central flow puzzle'. By consolidating diverse measurements into a coherent curve, these functions illuminate key factors influencing \(v_n(p_T,\text{cent})\), such as initial-state eccentricities (\(\varepsilon_{n}\)), dimensionless size (\(\mathbb{R}\)), radial flow, the medium's stopping power (\(\Hat{q}\)), and specific shear viscosity (\(\eta/s\)), along with viscous corrections (\(\delta_f\)). This comprehensive approach not only deepens our understanding of the transport properties and deformation of nuclei but also provides precise constraints for modeling the quark-gluon plasma.

\bibliography{refs}

\end{document}